# Efficient Diversification of Web Search Results

Gabriele Capannini
ISTI–CNR - Pisa, Italy
g.capannini@isti.cnr.it

Franco Maria Nardini
ISTI–CNR - Pisa, Italy
f.nardini@isti.cnr.it

Raffaele Perego
ISTI–CNR - Pisa, Italy
r.perego@isti.cnr.it

Fabrizio Silvestri
ISTI–CNR - Pisa, Italy
f.silvestri@isti.cnr.it

## ABSTRACT

In this paper we analyze the efficiency of various search results diversification methods. While efficacy of diversification approaches has been deeply investigated in the past, response time and scalability issues have been rarely addressed. A unified framework for studying performance and feasibility of result diversification solutions is thus proposed. First we define a new methodology for detecting when, and how, query results need to be diversified. To this purpose, we rely on the concept of "query refinement" to estimate the probability of a query to be *ambiguous*. Then, relying on this novel ambiguity detection method, we deploy and compare on a standard test set, three different diversification methods: IASelect, xQuAD, and OptSelect. While the first two are recent state-of-the-art proposals, the latter is an original algorithm introduced in this paper. We evaluate both the efficiency and the effectiveness of our approach against its competitors by using the standard TREC Web diversification track testbed. Results shown that OptSelect is able to run two orders of magnitude faster than the two other state-of-the-art approaches and to obtain comparable figures in diversification effectiveness.

## 1. INTRODUCTION

Web Search Engines (WSEs) are nowadays the most popular mean of interaction with the Web. Users interact with them by usually typing a few keywords representing their information need. Queries, however, are often ambiguous and have more than one possible interpretation [3, 19].

Consider, for example, the popular single-term query "apple". It might refer to Apple Corp., to the fruit, or to a tour operator which is very popular in the US. Without any further information that may help to disambiguate the user intent, search engines should produce a set of results possibly *covering* all (the majority of) the different interpretations of the query. To help users in finding the right information they are looking for, many different interfaces have been proposed to present search results. The first, and naïve, solution has been paging the ranked results list. Instead of presenting the whole list of results, the search engine presents results divided in pages (also known with the term SERP, i.e. Search Engine Result Page) containing ten results each. Some alternative interfaces have been proposed. *Results clustering* approaches, for instance, organize search results into folders that group similar items together [28, 9]. On the other side of the same coin, *results diversification* [1] follows an intermediate approach that aims at "packing" the highest possible number of diverse, relevant results within the first SERP.

Given that web search engines' mission of satisfying their users is of paramount importance, diversification of web search results is a hot research topic nowadays. Nevertheless, the majority of research efforts have been put on studying effective diversification methods able to satisfy web users. In this paper we take a different turn and consider the problem from the *efficiency* perspective. As Google's co-founder Larry Page declares[1]: "Speed is a major search priority, which is why in general we do not turn on new features if they will slow our services down." This paper, which extends a preliminary short version presented at WWW 2011 [16] is, to the best of our knowledge, the first work discussing efficiency in SERP diversification. The approach we follow to achieve an efficient and viable solution is based on analyzing query log information to infer knowledge about when diversification actions have to be taken, and what users expectations are. Indeed, the original contributions presented are several:

- We define a methodology for detecting when, and how, query results need to be diversified. We rely on the well-known concept of query refinement to estimate the probability of a query to be *ambiguous*. In addition, we show how to derive the most likely refinements, and how to use them to diversify the results.

- We define a novel *utility* measure to evaluate how useful is a result for a diversified result list.

- We propose OptSelect, an original algorithm allowing the diversification task to be accomplished effectively and very efficiently.

---

[1] http://www.google.com/corporate/tech.html

- Relying on our diversification framework, we deploy OptSelect and two other recent state-of-the-art diversification methods in order to evaluate on the standard TREC Web diversification track testbed both the efficiency and the effectiveness of our approach against its competitors.

The paper is organized as follows: Section 2 discusses related works. Section 3 presents a formalization of the problem, the specialization extraction method, and the algorithm we propose. Section 4 shows the efficiency of our solutions. Section 5 discusses some experimental results. In Section 6 we present our conclusions and we outline possible future work.

## 2. RELATED WORK

Result diversification has recently attracted a lot of interest. A very important pioneering work on diversification is [8]. In this paper, the authors present the Maximal Marginal Relevance (MMR) problem, and they show how a trade-off between novelty and relevance of search results can be made explicit through the use of two different functions, the first measuring the similarity among documents, and the other the similarity between documents and the query.

Zhai *et al.* [31] stated that in general it is not sufficient to return a set of relevant results as the correlation among the returned results is also very important. In a later work, Zhai *et al.* [30] formalize and propose a risk minimization approach that allow an arbitrary *loss* function over a set of returned documents to be defined. *Loss* functions aim at determining the *dissatisfaction* of the user with respect to a particular set of results. Such *loss* function depends on the language models rather than on categorical information about two documents [29].

Diversification has also been studied for purposes different from search engine result diversification. Ziegler *et al.* [32] study the diversification problem from a "recommendation" point of view. Radlinski *et al.* [20] propose a learning algorithm to compute an ordering of search results from a diverse set of orderings. Vee *et al.* [27] study the diversification problem in the context of structured databases with applications to online shopping. Clarke *et al.* [11] study diversification in question answering.

Agrawal *et al.* [1] present a systematic approach to diversify results that aims to minimize the risk of dissatisfaction of the web search engine users. Furthermore, authors generalize some classical IR metrics, including NDCG, MRR, and MAP, to explicitly account for the value of diversification. They show empirically that their algorithm scores higher in these generalized metrics compared to results produced by commercial search engines.

Gollapudi *et al.* [13] use the axiomatic approach to characterize and design diversification systems. They develop a set of axioms that a diversification system is expected to satisfy, and show that no diversification function can satisfy all these axioms simultaneously. Finally, they propose an evaluation methodology to characterize the objectives and the underlying axioms. They conduct a large scale evaluation based on data derived from Wikipedia and a product database.

Rafiei *et al.* [22] model the diversity problem as expectation maximization and study the challenges of estimating the model parameters and reaching an equilibrium. One model parameter, for example, is the correlation between pages which authors estimate using textual contents of pages and click data (when available). They conduct experiments on diversifying randomly selected queries from a query log and the queries chosen from the disambiguation topics of Wikipedia.

Clough *et al.* [12] examine user queries with respect to diversity. They analyze 14.9 million queries from the MSN query log by using two query log analysis techniques (click entropy and reformulated queries). Authors found that a broad range of query types may benefit from diversification. They also found that, although there is a correlation between word ambiguity and the need for diversity, the range of results users may wish to see for an ambiguous query stretches well beyond traditional notions of word sense.

Santos *et al.* [25, 24, 26] introduce a novel probabilistic framework (xQuAD) for Web search result diversification, which explicitly accounts for the various aspects associated to an underspecified query. In particular, they diversify a document ranking by estimating how well a given document satisfies each uncovered aspect and the extent to which different aspects are satisfied by the ranking as a whole. Authors evaluate the xQuAD framework in the context of the diversity task of the TREC 2009 Web track. They exploit query reformulations provided by three major WSEs to uncover different query aspects.

Similarly to [24], we exploit related queries as a mean of achieving diversification of query results. Nevertheless, our approach is very different from the above two. In [24], the authors exploit query reformulations provided by commercial Web search engines. All these reformulations are then taken into account during query processing in order to achieve a result set covering all the facets. In our case, the different meanings and facets of queries are instead disclosed by analyzing user behaviors recorded in query logs. During the analysis also the popularity of the different specializations is derived. Popularity distribution is then used to maximize the "usefulness" of the final set of documents returned. An approach orthogonal to our is instead investigated by Radlinski and Dumais in [19], where the problem of generating queries that can yield to a more diverse results set is studied starting from the observation that the top-$k$ results retrieved for a query might not contain representative documents for all of its interpretations.

## 3. DIVERSIFYING USING QUERY LOG

Users generally query a search engine by submitting a sequence of requests. Splitting the chronologically ordered sequence of queries submitted by a given user into *sessions*, is a challenging research topic [18, 4, 15]. Since our approach exploits user session information, but session splitting methodologies are out of the scope of this paper, we resort to adopt a state-of-the-art technique based on Query-Flow Graph [5, 6]. It consists of building a Markov Chain model of the query log and subsequently finding paths in the graph which are more likely to be followed by random surfers. As a result, by processing a query log $Q$ we obtain the set of logical user sessions exploited by our result diversification solution. Both the query topics possibly benefiting from diversification, and the probability of each distinct specialization among the spectrum of possibilities, are in fact mined from logs storing historical information about the interaction of users with the WSE.

As an example, let us assume that in a given query log the queries *leopard mac OS X*, *leopard tank*, and *leopard pictures*, are three specializations of query *leopard* that commonly occur in logical query sessions. The presence of the same query refinements in several sessions issued by different users gives us evidence that a query is ambiguous, while the relative popularity of its specializations allow us to compute the probabilities of the different meanings. On the basis of this information learned from historical data, once a query $q$ is encountered by the WSE, we: (a) check if $q$ is ambiguous or faceted, and if so, (b) exploit the knowledge about the different specializations of $q$ submitted in the past to retrieve documents relevant for all of them. Finally, (c) use the relative frequencies of these specializations to build a final result set that maximize the probability of satisfying the user. In the following, we describe more precisely how ambiguous/faceted queries are detected and managed.

## 3.1 Mining Specializations from Query Logs

We assume that a query log $Q$ is composed by a set of records $\langle q_i, u_i, t_i, V_i, C_i \rangle$ storing, for each submitted query $q_i$: (i) the anonymized user $u_i$; (ii) the timestamp $t_i$ at which $u_i$ issued $q_i$; (iii) the set $V_i$ of URLs of documents returned as top-$k$ results of the query, and, (iv), the set $C_i$ of URLs corresponding to results clicked by $u_i$. Let $q$ and $q'$ be two queries submitted by the same user during the same logical session recorded in $Q$. We adopt the terminology proposed in [6], and we say that a query $q'$ is a "specialization" of $q$ if the user information need is stated more precisely in $q'$ than in $q$ (i.e., $q'$ is more specific than $q$). Let us call $S_q$ the set of specializations of an ambiguous/faceted query $q$ mined from the query log.

Given the above generic definition, any algorithm that exploits the knowledge present in query log sessions to provide users with useful suggestions of related queries, can be easily adapted to the purpose of devising specializations of submitted queries. Given the popularity function $f()$ that computes the frequency of a query topic in $Q$, and a query recommendation algorithm $\mathcal{A}$ trained with $Q$, Algorithm 1 can be used to detect efficiently and effectively queries that can benefit from result diversification, and to compute for them the set of most common specializations along with their probabilities.

---

**Algorithm 1** AmbiguousQueryDetect($q, \mathcal{A}, f(), s$)

/\* given the submitted query $q$, a query recommendation algorithm $\mathcal{A}$, and an integer $s$ compute the set $\widehat{S}_q \subseteq Q$ of possible specializations of $q$ \*/
1. $\widehat{S}_q \leftarrow \mathcal{A}(q)$;
/\* select from $\widehat{S}_q$ the most popular specializations \*/
2. $S_q \leftarrow \{q' \in \widehat{S}_q | f(q') \geq \frac{f(q)}{s}\}$;
3. **If** $|S_q| \geq 2$ **Then Return** ($S_q$); **Else Return** ($\emptyset$);

---

In this work, we experimented the use of a very efficient query recommendation algorithm [7] for computing the possible specializations of queries. The algorithm used learns the suggestion model from the query log, and returns as related specializations, only queries that are present in $Q$, and for which related probabilities can be, thus, easily computed. Note that any other approach for deriving user intents from query logs, (as an example, [21, 23]), could be used and easily integrated in our diversification framework.

DEFINITION 1 (PROBABILITY OF SPECIALIZATION). *Let $\widehat{Q} = \{q \in Q, \ s.t. \ |S_q| > 1\}$ be the set of ambiguous queries in $Q$, and $P(q'|q)$ the probability for $q \in \widehat{Q}$ to be specialized from $q' \in S_q$.*

We assume that the distribution underlying the possible specialization of an ambiguous query is known and complete, i.e., $\sum_{q' \in S_q} P(q'|q) = 1$, and $P(q'|q) = 0, \forall q' \notin S_q, \forall q \in \widehat{Q}$. To our purposes these probability distributions are simply estimated by dividing the frequency returned by Algorithm 1 using the following formula:

$$P(q'|q) = f(q') / \sum_{q' \in S_q} f(q')$$

Obviously, query logs can not give the complete knowledge about all the possible specializations for a given ambiguous query, but we can expect that the most popular interpretations are present in a large query log covering a long time period. Now, let us give some additional assumptions and notations.

$\mathcal{D}$ is the collection of documents indexed by the search engine which returns, for each submitted query $q$, an ordered list $R_q$ of documents. The rank of document $d \in \mathcal{D}$ within $R_q$ is indicated with $rank(d, R_q)$.

Moreover, let $d_1$ and $d_2$ be two documents of $\mathcal{D}$, and $\delta : \mathcal{D} \times \mathcal{D} \to [0,1]$ a distance function having the non-negative, and symmetric properties, i.e. (i) $\delta(d_1, d_2) = 0$ iff $d_1 = d_2$, and (ii) $\delta(d_1, d_2) = \delta(d_2, d_1)$.

DEFINITION 2 (RESULTS' UTILITY). *The utility of a result $d \in R_q$ for a specialization $q'$ is defined as:*

$$U(d|R_{q'}) = \sum_{d' \in R_{q'}} \frac{1 - \delta(d, d')}{rank(d', R_{q'})}. \tag{1}$$

*where $R_{q'}$ is the list of results that the search engine returned for specialized query $q'$.*

Such utility represents how good $d \in R_q$ is for satisfying a user intent that is better represented by specialization $q'$. The intuition for $U$ is that a result $d \in R_q$ is more useful for specialization $q'$ if it is very similar to a highly ranked item contained in the results list $R_{q'}$.

The utility function specified in Equation (1) uses the following function to measure the distance between two documents:

$$\delta(d_1, d_2) = 1 - cosine(d_1, d_2) \tag{2}$$

where $cosine(d_1, d_2)$ is the cosine similarity between the two documents.

In the methods presented in the following, we use a normalized version of results' utility, $\widetilde{U}(d|R_{q'})$, which is defined as the normalization of $U(d|R_{q'})$ in the $[0,1]$ interval. The normalization factor is computed by assuming that, in the optimal case, result $d$ is at distance $\delta(d, \cdot) = 0$. In this case, the utility function is equal to

$$\sum_{d' \in R_{q'}} \frac{1}{rank(d', R_{q'})} = \sum_{i=1}^{|R_{q'}|} \frac{1}{i} = H_{|R_{q'}|}$$

where $H_{|R_{q'}|}$ is the $|R_{q'}|$-th harmonic number. Therefore,

$$\widetilde{U}(d|R_{q'}) = \frac{U(d|R_{q'})}{H_{|R_{q'}|}}$$

Using the above definitions, we can now define three different query-logs-based approaches to diversification. The first two methods are adaptations of the Agrawal *et al.* [1] algorithm, and the Santos's *et al.* xQuAD framework [24]. The last one refers to our novel formulation detailed in Section 3.1.3.

### 3.1.1 The QL_DIVERSIFY($k$) Problem

In a recent paper, Agrawal *et al.* [1] defined the DIVERSIFY($k$) problem, a covering-like problem aimed to include the maximum number of possible "categories" into the list of $k$ results that are returned in response to a user's query.

We briefly recall the definition of the problem, as stated in the original paper [1]:

DIVERSIFY($k$): Given query $q$, a set of documents $R_q$, a probability distribution of categories for the query $P(c|q)$, the quality values of the documents $V(d|q,c)$, $\forall d \in \mathcal{D}$ and an integer $k$. Find a set of documents $S \subseteq R_q$ with $|S| = k$ that maximizes

$$P(S|q) = \sum_c P(c|q) \left( 1 - \prod_{d \in S} (1 - V(d|q,c)) \right) \quad (3)$$

Equation (3) uses two concepts similar to those we have already introduced: the probability of a query to be part of a category is very similar to our concept of probability of specialization (see Definition 1), while quality value $V(d|q,c)$ resembles $\widetilde{U}(d|R_{q'})$.

It is possible to see the set of possible specializations $S_q$ as the set of possible categories for $q$ mined from a query log. The utility, in this case, can be seen as the utility of selecting resulting document $d$ for category/specialization $q'$. Thus, the problem becomes choosing a subset $S$ of $R_q$ with $|S| = k$ that maximizes:

$$P(S|q) = \sum_{q' \in S_q} P(q'|q) \left( 1 - \prod_{d \in S} \left( 1 - \widetilde{U}(d|R_{q'}) \right) \right) \quad (4)$$

We call this problem QL_DIVERSIFY($k$) to differentiate it from the original Agrawal *et al.* formulation [1].

### 3.1.2 The xQuAD_DIVERSIFY($k$) Problem

In [24], Santos *et al.* propose a probabilistic framework called xQuAD. Compared to [1], the proposed framework extends the measure with which documents produced for ambiguous query $q$ are iteratively selected. To this aim, xQuAD evaluates also the initial ranking of such documents for $q$. Formally, the problem is the following:

xQuAD_DIVERSIFY($k$): Given a query $q$, a set of ranked documents $R_q$ retrieved for $q$, a mixing parameter $\lambda \in [0, 1]$, two probability distributions $P(d|q)$ and $P(d, \bar{S}|q)$ measuring, respectively, the likelihood of document $d$ being observed given $q$, and the likelihood of observing $d$ but not the documents in the solution $S$. Find a set of documents $S \subseteq R_q$ with $|S| = k$ that maximizes for each $d \in S$

$$(1 - \lambda) \cdot P(d|q) + \lambda \cdot P(d, \bar{S}|q) \quad (5)$$

xQuAD is a greedy algorithm that iteratively selects a new document, and pushes it into the current solution. The selection process consists in choosing each time the document $d^* \in \bar{S} = R \setminus S$ that maximizes Equation (5). Such formula combines two probabilities: the first evaluates the *relevance* of a document $d$ as the expectation for $d$ to be observed given the query $q$, namely $P(d|q)$. The second probability measures the *diversity* of a candidate document $d$ as the product of two components (see Equation (6)). The first component is, thus, the relevance of $d$ with respect to a set of specializations $S_q$, and it is obtained by multiplying the likelihood of a specialization $q'$ by the likelihood of $d$ considering a particular specialization $q'$. Furthermore, the second component estimates the coverage degree of the current solution $S$ with respect to each specialization $q'$.

$$P(d, \bar{S}|q) = \sum_{q' \in S_q} \left[ P(q'|q) P(d|q') \prod_{d_j \in S} 1 - P(d_j|q') \right] \quad (6)$$

As for the Agrawal's formulation, $P(d_j|q')$ can be seen as the utility of selecting resulting document $d_j$ for specialization $q'$. Thus, we measure $P(d_j|q')$ using $\widetilde{U}(d|R_{q'})$. Similarly to [1], at each step, the algorithm updates the coverage degree of solution $S$ for each candidate document, then it scans $R \setminus S$ in order to choose the best document.

### 3.1.3 The MaxUtility_DIVERSIFY($k$) Problem

The problem addressed in the Agrawal's paper, is actually the maximization of the *weighted* coverage of the categories with pertinent results. The objective function does not consider directly the number of categories covered by the result set; it might be the case that even if the categories are less than $|S_q|$, some of these will not be covered by the results set. This may happen because the objective function considers explicitly how much a document satisfies a given category. Hence, if a category that is a dominant interpretation of the query $q$ is not covered adequately, more documents related to such category will be selected, possibly at the expense of other categories.

We believe, instead, that it is possible to maximize the sum of the various utilities for the chosen subset $S$ of documents by guaranteeing that query specializations are covered proportionally to the associated probabilities $P(q'|q)$. Motivated by the above observation, we define the following problem.

MAXUTILITY_DIVERSIFY($k$): Given a query $q$, the set $R_q$ of results for $q$, two probability distributions $P(d|q)$ and $P(q'|q) \forall q' \in S_q$ measuring, respectively, the likelihood of document $d$ being observed given $q$, and the likelihood of having $q'$ as a specialization of $q$, the utilities $\widetilde{U}(d|R_{q'})$ of documents, a mixing parameter $\lambda \in [0, 1]$, and an integer $k$. Find a set of documents $S \subseteq R_q$ with $|S| = k$ that maximizes

$$\widetilde{U}(S|q) = \sum_{d \in S} \sum_{q' \in S_q} (1 - \lambda) P(d|q) + \lambda P(q'|q) \widetilde{U}(d|R_{q'}) \quad (7)$$

with the constraints that every specialization is covered proportionally to its probability. Formally, let $R_q \bowtie q' = \{d \in R_q | U(d|R_{q'}) > 0\}$. We require that for each $q' \in S_q$, $|R_q \bowtie q'| \geq \lfloor k \cdot P(q'|q) \rfloor$.

Our technique aims at selecting from $R_q$ the $k$ results that maximize the overall utility of the list of results. When $|S_q| \leq k$ the results are in someway split into $|S_q|$ subsets each one covering a distinct specializations. The more popular a specialization, the greater the number of results relevant for it. Obviously, if $|S_q| > k$ we select from $S_q$ the $k$ specializations with the largest probabilities.

## 4. EFFICIENCY EVALUATION

Efficiency of diversification algorithms is an important issue to study. Even the best diversification algorithm can be useless if its high computational cost forbids its actual use in a real-world IR system. In the following discussion, IASelect is the greedy algorithm used to approximate QL_DIVERSIFY($k$), xQuAD refers to the greedy algorithm used to approximate xQuAD_DIVERSIFY($k$), and eventually OptSelect is our algorithm solving the MAXUTILITY_DIVERSIFY($k$) problem. We consider diversification to be done on a set of $|R_q| = n$ results returned by the baseline retrieval algorithm. Furthermore, we consider $|S_q|$, i.e. the number of specifications for a query $q$ to be a constant (indeed, it is usually a small value depending on $q$).

**IASelect.** As shown by Agrawal *et al.* the DIVERSIFY($k$) problem, and thus also the QL_DIVERSIFY($k$) problem, is NP-Hard. Since the problem's objective function is *submodular*, an opportune greedy algorithm yields to a solution whose value is smaller than $(1 - 1/e)$ times the optimal one [17]. The greedy algorithm consists in adding to the results set the documents giving the largest marginal increase to the objective function. Since there is an insertion operation for each result needed in the final result set, the algorithm performs $k$ insertions. For each insertion the algorithm searches for the document with the largest marginal utility that has not yet been selected. Since marginal utility is computed for each candidate document in terms of the current solution and each specialization, its value must be updated at each insertion. Hence, the computational cost of the procedure is linear in the number of categories/specializations multiplied by the number $n$ of candidate documents. Thus, the solution proposed has a cost $C_{\text{I}}(n,k) = \sum_{i=1}^{k} [\ |S_q| \cdot (n-i)\ ] = k|S_q| \left(n - \frac{k+1}{2}\right) = O(nk)$.

**xQuAD.** It is a greedy algorithm that iteratively selects a new document, and pushes it into the current solution. The selection process consists in choosing each time the document $d^* \in R \setminus S$ that maximizes Equation (5). As specified in Section 3.1.2, such formula combines the probability for a document $d$ of being relevant for a query $q$, i.e., $P(d|q)$ and the *diversity* of a candidate document $d$, respectively.

Similarly to the solution proposed in [1], at each step, the algorithm updates the coverage degree of solution $S$ for each candidate document, then it scans $R \setminus S$ in order to choose the best document. The procedure is linear in the number of items in $S_q$ multiplied by the number of documents in $R \setminus S$. Since the selection is performed $k$ times, the final computational cost is given by $C_{\text{X}}(n,k) = \sum_{i=1}^{k} [\ |S_q| \cdot (n-i)\ ]$. As for the Agrawal's solution, thus, $C_{\text{X}}(n,k) = O(nk)$.

**Optselect.** While QL_DIVERSIFY($k$) aims to maximize the probability of covering useful categories, the MAXUTILITY_DIVERSIFY($k$) aims to maximize directly the overall utility. This simple relaxation allows the problem to be simplified and solved *optimally* in a very simple and efficient way. Furthermore, the constraints bounding the minimum number of results tied to a given specialization, boost the quality of the final diversified result list, ensuring that the covered specializations reflect the most popular preferences expressed by users in the past.

Another important difference between Equation (7) and Equation (4) is that the latter needs to select, in advance, the subset $S$ of documents before computing the final score. In our case, instead, a simple arithmetic argument shows that:

$$\widetilde{U}(S|q) = \sum_{d \in S} \widetilde{U}(d|q) \qquad (8)$$

where $\widetilde{U}(d|q)$ is the overall utility of document $d$ for query $q$. This value is computed according to the following equation:

$$\widetilde{U}(d|q) = \sum_{q' \in S_q} (1-\lambda) P(d|q) + \lambda P(q'|q) \widetilde{U}(d|R_{q'}) \qquad (9)$$

By combining (8), and (9) we obtain:

$$\begin{aligned}\widetilde{U}(S|q) &= (1-\lambda)|S_q| \sum_{d \in S} P(d|q) + \\ &+ \lambda \sum_{q' \in S_q} P(q'|q) \sum_{d \in S} \widetilde{U}(d|R_{q'})\end{aligned}$$

Therefore, to maximize $\widetilde{U}(S|q)$ we simply resort to compute for each $d \in R_q$: i) the relevance of $d$ for the query $q$, ii) the utility of $d$ for specializations $q' \in S_q$ and, then, to *select* the top-$k$ highest ranked documents. Obviously, we have to carefully select results to be included in the final list in order to avoid choosing results that are relevant only for a single specialization. For this reason we use a collection of $|S_q|$ heaps each of those keeps the top $\lfloor k \cdot P(q'|q) \rfloor + 1$ most useful documents for that specialization. Algorithm 2 in Appendix A returns the set $S$ maximizing the objective function in Equation (7). Moreover, the running time of the algorithm is linear in the size of the documents considered. Indeed, all the heap operations are carried out on data structures having a constant size bounded by $k$.

Similarly to the other two solutions discussed, the proposed solution is computed by using a greedy algorithm. OptSelect is however computationally less expensive than its competitors. The main reason is that for each inserted element, it does not recompute the marginal utility of the remaining documents w.r.t. all the specializations. The main computational cost is given by the procedure which tries to add elements to each heap related to a specialization in $S_q$. Since each heap is of at most $k$ positions, each insertion has cost $\log_2 k$, and globally the algorithm costs $C_{\text{O}}(n,k) = n|S_q| \log_2 k = O(n \log_2 k)$.

Table 1 reports and compares the theoretical complexity of the three considered methods. Our newly proposed OptSelect algorithm is faster than the previously proposed ones.

**Empirical efficiency evaluation.** In addition to the theoretical considerations above, we also conducted tests in the TREC 2009 Web track's Diversity Task framework to empirically compare the efficiency of the three solutions proposed. In particular, we measured the time required by OptSelect, xQuAD and IASelect to diversify the list of retrieved documents. All the tests were conducted on a Intel Core 2

| Algorithm | Complexity |
|---|---|
| IASelect | $O(nk)$ |
| xQuAD | $O(nk)$ |
| OptSelect | $O(n\log_2 k)$ |

Table 1: Time complexity of the three algorithms considered.

Quad PC with 8Gb of RAM and Ubuntu Linux 9.10 (kernel 2.6.31-22).

Table 2 reports the average time required by the three algorithms to diversify the initial set of documents for the 50 queries of the TREC 2009 Web Track's Diversity Task. We study the performance by varying both the number of documents which the diversified result set is chosen from ($|R_q|$), and the size of the returned list $S$ denoted by $k$ (i.e. $k = |S|$). The results show that, for each value of $k$, the execution time of all the tested methods is linear by varying the size of $R_q$. The only difference among these trends is in favor of OptSelect which slope is lower than its competitors. By varying, instead, the value of $k$, the execution times follow the complexities resumed in Table 1. The most remarkable result is that, increasing the number of documents returned, OptSelect outperforms xQuAD and IASelect in all the conducted tests. In particular, OptSelect is two orders of magnitude faster than its competitors.

| $|R_q|$ | $k$ | | | | |
|---|---|---|---|---|---|
| | 10 | 50 | 100 | 500 | 1000 |
| OptSelect | | | | | |
| 1,000 | 0.34 | 0.58 | 0.66 | 0.89 | 0.98 |
| 10,000 | 1.36 | 2.13 | 2.46 | 3.32 | 3.57 |
| 100,000 | 4.81 | 8.32 | 9.57 | 12.94 | 13.92 |
| xQuAD | | | | | |
| 1,000 | 0.43 | 1.64 | 3.31 | 14.82 | 30.18 |
| 10,000 | 3.27 | 16.69 | 32.22 | 148.41 | 298.63 |
| 100,000 | 36.27 | 143.67 | 285.69 | 1,425.82 | 2,849.83 |
| IASelect | | | | | |
| 1,000 | 0.57 | 1.68 | 3.92 | 20.81 | 39.82 |
| 10,000 | 4.23 | 23.03 | 40.82 | 203.11 | 409.43 |
| 100,000 | 48.04 | 205.46 | 408.61 | 2,039.22 | 4,071.81 |

Table 2: Execution time (in msec.) of OptSelect, xQuAD, and IASelect by varying both the size of the initial set of documents to diversify ($|R_q|$), and the size of the diversified result set ($k = |S|$).

## 4.1 Feasibility of the Diversification Solution

Something worth to be discussed is the feasibility of our diversification solution. Differently from other approaches, our solution does not require any pre-existing taxonomy (or classification model) built in advance. The only information we need are: the ambiguous queries, the list of their possible specializations mined from a long-term query log, the probabilities associated with such specializations, and the sets $R_{q'}$ of documents highly relevant for each specialization. It is worth noting that the number of documents highly relevant for each specialization that need to be maintained is very small compared to the set of documents $R_q$ to be re-ranked on the basis of the specializations, i.e., $|R_{q'}| \ll |R_q|$. A *back-of-the-envelope* computation highlights the small footprint of the data structures needed to actually implement our method. Given the ambiguous query $\widehat{q}$ having the largest number $|S_{\widehat{q}}|$ of specializations, we have to store $|R_{\widehat{q}'}|$ documents for each one of the specializations $\widehat{q}'$. Let $L$ be the average length in bytes of these documents. Actually only short summaries, and not whole documents, can be used without significative loss in the precision of our method. Resuming, storing $N$ ambiguous query along with the data needed to assess the similarity among results lists incurs in a maximal memory occupancy of $N \cdot |S_{\widehat{q}}| \cdot |R_{\widehat{q}'}| \cdot L$ bytes.

## 5. TESTING EFFECTIVENESS

We conducted our experiments to measure the effectiveness of the three methods in the context of the diversity task of the TREC 2009 Web track [10]. The goal of this task is to produce a ranking of documents for a given query that maximizes the coverage of the possible aspects underlying this query, while reducing its overall redundancy with respect to the covered aspects. In our experiments, we used ClueWeb-B, the subset of the TREC ClueWeb09 dataset[2] and two query logs (AOL and MSN). Both ClueWeb-B and the two query logs used are described in Appendix B. The query associated with each topic of the TREC 2009 Web track was used as initial ambiguous/faceted query.

The two query logs were first preprocessed in order to devise the logical user sessions as described in Section 3. Moreover the sessions obtained were used to build the model for the recommendation algorithm described in [7]. Given a query $q$, such algorithm was used to compute efficiently the set and the associated probabilities of its popular specializations $S_q$ (see Algorithm 1).

The results obtained for the diversity task of the TREC 2009 Web track are evaluated according to the two official metrics: $\alpha$-NDCG and IA-P. The $\alpha$-normalized discounted cumulative gain ($\alpha$-NDCG [11]) metric balances relevance and diversity through the tuning parameter $\alpha$. The larger the value of $\alpha$, the more diversity is rewarded. In contrast, when $\alpha = 0$, only relevance is rewarded, and this metric is equivalent to the traditional NDCG [14]. Moreover, we used the intent-aware precision (IA-P [1]) metric, which extends the traditional notion of precision in order to account for the possible aspects underlying a query and their relative importance. In our evaluation, both $\alpha$-NDCG and IA-P are reported at five different rank cutoffs: 5, 10, 20, 100, and 1000. While the first four cutoffs focus on the evaluation at early ranks which are very important in a web context, the last cutoff gives the value of the two metrics for all the set of results. Both $\alpha$-NDCG and IA-P are computed following the standard practice in the TREC 2009 Web-Track's Diversity Task [10]. In particular, $\alpha$-NDCG is computed with $\alpha = 0.5$, in order to give an equal weight to relevance and diversity.

An ad-hoc modified version of the Terrier[3] IR platform was used for both indexing and retrieval. We extended Terrier in order to obtain short summaries of retrieved documents, which are used as document surrogates in our diversification algorithm. We used Porter's stemmer and standard English stopword removal for producing the ClueWeb-B index. We evaluate the effectiveness of our method in diversifying the results retrieved using a probabilistic document weighting model: DPH Divergence From Randomness (DFR) model [2].

---

[2] http://boston.lti.cs.cmu.edu/Data/clueweb09/

[3] http://www.terrier.org

Table 3 shows the results of the tests conducted with the DPH baseline (no diversification), i) our OptSelect, ii) Agrawal's IASelect, and iii) the xQuAD framework. We set $|R_{q'}| = 20$, $k = 1000$, and $|R_q| = 25,000$. Furthermore, xQuAD and OptSelect use a value for parameter $\lambda$ equal to 0.15 (the value maximizing $\alpha$-NDCG@20 in [24]). We applied the utility function in (1) to the snippets returned by the Terrier search engine instead of applying it to the whole documents, and we forced its returning value to be 0 when it is below a given threshold $c$. Nine different values of the utility threshold $c$ were tested. The specializations and the associated probabilities were obtained in all the cases by using the previously described approach [7].

The results reported in the Table show that OptSelect and xQuAD behave similarly, while IASelect performs always worse. OptSelect shows good performance for small values of $c$, in particular for $c \in \{0, 0.05\}$. For both the two values of the threshold, OptSelect obtains very good $\alpha$-NDCG performance and the best IA-P values. A deeper analysis of Table 3 shows that OptSelect obtains better results than the other two methods in terms of IA-P@5 for $c = 0.05$. The best $\alpha$-NDCG performances for OptSelect are instead obtained for $c = 0.20$. For this value of the threshold, OptSelect shows a good trade-off between $\alpha$-NDCG and IA-P, in particular for short results' lists (@5, @10, @20). However, none of these differences can be classified as statistically significant according to the Wilcoxon signed-rank test at 0.05 level of significance. Increasing the value of the threshold $c$, effectiveness starts to degrade. In fact, for $c \geq 0.75$ all the algorithms perform basically as the DPH baseline.

The xQuAD framework obtains good $\alpha$-NDCG and IA-P performance for $c = 0.05$. xQuAD performs well also for $c = 0.20$. Note that our formulation of the xQuAD framework performs better than reported in the original paper by Santos *et al.* [24]. Essentially, this behavior could be explained by the following two reasons: i) our method for measuring the "diversity" of a document based on Equation (1) is superior to the one used in [24], ii) our method for deriving specializations, and their associated probabilities is able to carry out more accurate results. We leave this analysis to a future work. By comparing OptSelect ($c = 0.20$) and xQuAD ($c = 0.05$), we highlight better performances for OptSelect in terms of IA-P@5, and IA-P@20, while the xQuAD framework slightly outperforms OptSelect for $\alpha$-NDCG, with an exception for $\alpha$-NDCG@5 where the two methods behave similarly.

Agrawal's IASelect shows its best performances when the threshold $c$ is not used. However, it never outperforms OptSelect and xQuAD. Both $\alpha$-NDCG and IA-P values improve over the DPH baseline but are always remarkably lower than the best values obtained using OptSelect and QuAD.

## 6. CONCLUSIONS AND FUTURE WORK

We studied the problem of diversifying search results by exploiting the knowledge derived from query logs. We presented a general framework for query result diversification comprising: (*i*) an efficient and effective methodology, based on state-of-the-art query recommendation algorithms, to detect ambiguous queries that would benefit from diversification, and to devise all the possible common specializations to be included in the diversified list of results along with their probability distribution; (*ii*) OptSelect: a new diversification algorithm which re-ranks the original results list on the basis of the mined specializations.

A novel formulation of the problem has been proposed and motivated. It allows the diversification problem to be modeled as a maximization problem. The approach is evaluated by using the metrics and the datasets provided for the TREC 2009 Web Track's Diversity Task. Our experimental results show that our approach is both efficient and effective. In terms of efficiency, our approach performs two orders of magnitude faster than its competitors and it remarkably outperforms its competitors in all the tests.

In terms of effectiveness, our approach outperforms the Agrawal's IASelect, and it shows the best results in terms of IA-P [1]. It produces also results that are comparable with the xQuAD framework in terms of $\alpha$-NDCG [11].

Future work will regard: i) the exploitation of users' search history for personalizing result diversification, ii) the use of click-through data to improve our effectiveness results, and iii) the study of a search architecture performing the diversification task in parallel with the document scoring phase.

## 7. ACKNOWLEDGMENTS

We acknowledge the partial support of S-CUBE (EU-FP7-215483) and ASSETS (CIP-ICT-PSP-250527) projects.

## 8. REFERENCES

[1] R. Agrawal, S. Gollapudi, A. Halverson, and S. Ieong. Diversifying search results. In *Proc. WSDM'09*, pages 5–14. ACM, 2009.
[2] G. Amati, E. Ambrosi, M. Bianchi, C. Gaibisso, and G. Gambosi. FUB, IASI-CNR and university of Tor Vergata at trec 2007 blog track. In *Proc. TREC*, volume Special Publication 500-274, 2007.
[3] A. Anagnostopoulos, A. Z. Broder, and D. Carmel. Sampling search-engine results. In *Proc. WWW'05*, pages 245–256. ACM, 2005.
[4] R. Baeza-Yates. Graphs from search engine queries. In *Proc. SOFSEM'07*, pages 1–8, Harrachov, CZ, 2007.
[5] P. Boldi, F. Bonchi, C. Castillo, D. Donato, A. Gionis, and S. Vigna. The query-flow graph: model and applications. In *In Proc. CIKM'08*, pages 609–618. ACM, 2008.
[6] P. Boldi, F. Bonchi, C. Castillo, and S. Vigna. From 'dango' to 'japanese cakes': Query reformulation models and patterns. In *Proc. WI'09*, pages 183–190. IEEE CS Press, 2009.
[7] D. Broccolo, L. Marcon, F. M. Nardini, R. Perego, and F. Silvestri. An efficient algorithm to generate search shortcuts. Technical Report N. /cnr.isti/2010-TR-017, CNR ISTI Pisa Italy, 2010.
[8] J. Carbonell and J. Goldstein. The use of MMR, diversity-based reranking for reordering documents and producing summaries. In *Proc. SIGIR'98*, pages 335–336. ACM, 1998.
[9] C. Carpineto, S. Osiński, G. Romano, and D. Weiss. A survey of web clustering engines. *ACM Comput. Surv.*, 41(3):1–38, 2009.
[10] C. Clarke, N. Craswell, and I. Soboroff. Preliminary report on the TREC 2009 Web track. 2009.
[11] C. L. Clarke, M. Kolla, G. V. Cormack, O. Vechtomova, A. Ashkan, S. Büttcher, and I. MacKinnon. Novelty and diversity in information retrieval evaluation. In *Proc. SIGIR'08*, pages 659–666. ACM, 2008.
[12] P. Clough, M. Sanderson, M. Abouammoh, S. Navarro, and M. Paramita. Multiple approaches to analysing query diversity. In *Proc SIGIR'09*, pages 734–735. ACM, 2009.
[13] S. Gollapudi and A. Sharma. An axiomatic approach for result diversification. In *Proc. WWW'09*, pages 381–390. ACM, 2009.

|  | $c$ | $\alpha$-NDCG | | | | | IA-P | | | | |
|---|---|---|---|---|---|---|---|---|---|---|---|
|  |  | @5 | @10 | @20 | @100 | @1000 | @5 | @10 | @20 | @100 | @1000 |
| DPH Baseline | - | 0.190 | 0.212 | 0.240 | 0.303 | 0.303 | 0.092 | 0.093 | 0.088 | 0.058 | 0.006 |
| OptSelect | 0 | **0.213** | 0.227 | 0.255 | 0.318 | 0.352 | 0.111 | 0.100 | **0.092** | 0.061 | 0.012 |
|  | 0.05 | **0.213** | 0.228 | 0.256 | 0.319 | 0.352 | **0.112** | **0.101** | 0.091 | 0.061 | 0.012 |
|  | 0.10 | 0.195 | 0.220 | 0.246 | 0.312 | 0.343 | 0.102 | 0.097 | 0.090 | **0.062** | 0.012 |
|  | 0.15 | 0.190 | 0.216 | 0.246 | 0.305 | 0.341 | 0.101 | 0.098 | 0.090 | 0.061 | 0.012 |
|  | 0.20 | **0.214** | **0.241** | **0.262** | **0.324** | **0.359** | 0.110 | 0.101 | 0.090 | 0.060 | 0.012 |
|  | 0.25 | 0.190 | 0.213 | 0.238 | 0.305 | 0.339 | 0.095 | 0.098 | 0.087 | 0.058 | 0.012 |
|  | 0.35 | 0.186 | 0.206 | 0.235 | 0.302 | 0.335 | 0.089 | 0.090 | 0.086 | 0.058 | 0.012 |
|  | 0.50 | 0.186 | 0.208 | 0.236 | 0.300 | 0.334 | 0.091 | 0.091 | 0.087 | 0.058 | 0.012 |
|  | 0.75 | 0.190 | 0.212 | 0.240 | 0.303 | 0.337 | 0.092 | 0.093 | 0.088 | 0.058 | 0.012 |
| xQuAD | 0 | 0.211 | 0.241 | 0.260 | 0.320 | 0.354 | 0.103 | 0.102 | 0.090 | 0.058 | 0.012 |
|  | 0.05 | **0.214** | **0.242** | 0.260 | **0.323** | **0.355** | **0.108** | **0.103** | 0.089 | 0.058 | 0.012 |
|  | 0.10 | 0.193 | 0.226 | 0.249 | 0.308 | 0.341 | 0.101 | 0.101 | 0.090 | 0.058 | 0.012 |
|  | 0.15 | 0.200 | 0.227 | 0.253 | 0.315 | 0.348 | 0.099 | 0.095 | 0.087 | 0.058 | 0.012 |
|  | 0.20 | 0.204 | 0.234 | **0.262** | 0.321 | 0.354 | 0.096 | 0.099 | 0.087 | 0.058 | 0.012 |
|  | 0.25 | 0.181 | 0.211 | 0.236 | 0.303 | 0.336 | 0.090 | 0.095 | 0.085 | 0.058 | 0.012 |
|  | 0.35 | 0.185 | 0.209 | 0.239 | 0.302 | 0.335 | 0.091 | 0.092 | 0.088 | 0.058 | 0.012 |
|  | 0.50 | 0.190 | 0.212 | 0.240 | 0.303 | 0.336 | 0.092 | 0.093 | 0.087 | 0.058 | 0.012 |
|  | 0.75 | 0.190 | 0.212 | 0.240 | 0.303 | 0.337 | 0.092 | 0.093 | 0.088 | 0.058 | 0.012 |
| IASelect | 0 | **0.206** | **0.215** | **0.245** | **0.302** | 0.334 | 0.097 | 0.089 | 0.079 | 0.044 | 0.009 |
|  | 0.05 | 0.205 | 0.214 | 0.243 | 0.299 | 0.330 | **0.098** | 0.090 | 0.078 | 0.044 | 0.009 |
|  | 0.10 | 0.193 | 0.200 | 0.227 | 0.279 | 0.309 | **0.098** | 0.088 | 0.075 | 0.039 | 0.008 |
|  | 0.15 | 0.169 | 0.185 | 0.207 | 0.259 | 0.288 | 0.089 | 0.078 | 0.064 | 0.039 | 0.008 |
|  | 0.20 | 0.182 | 0.197 | 0.229 | 0.284 | 0.314 | 0.085 | 0.074 | 0.067 | 0.046 | 0.009 |
|  | 0.25 | 0.198 | 0.214 | 0.243 | 0.301 | 0.332 | 0.092 | 0.083 | 0.076 | 0.052 | 0.011 |
|  | 0.35 | 0.192 | 0.208 | 0.241 | 0.299 | 0.332 | 0.095 | **0.093** | 0.087 | 0.057 | 0.012 |
|  | 0.50 | 0.192 | 0.214 | 0.243 | 0.306 | **0.338** | 0.093 | 0.091 | 0.087 | **0.058** | 0.012 |
|  | 0.75 | 0.190 | 0.212 | 0.240 | 0.303 | 0.337 | 0.092 | **0.093** | **0.088** | **0.058** | 0.012 |

Table 3: Values of $\alpha$-NDCG, and IA-P for OptSelect, xQuAD, and IASelect by varying the threshold $c$.

[14] K. Järvelin and J. Kekäläinen. Cumulated gain-based evaluation of ir techniques. *ACM Trans. Inf. Syst.*, 20(4):422–446, 2002.
[15] R. Jones and K. L. Klinkner. Beyond the session timeout: automatic hierarchical segmentation of search topics in query logs. In *Proc. CIKM'08*, pages 699–708. ACM, 2008.
[16] F. M. Nardini, G. Capannini, R. Perego, and F. Silvestri. Efficient diversification of search results using query logs. In *Proc. WWW'11*, page to appear, New York, NY, USA, 2011. ACM.
[17] G. Nemhauser, L. Wolsey, and M. Fisher. An analysis of the approximations for maximizing submodular set functions. *Mathematical Programming*, 14:265–294, 1978.
[18] B. Piwowarski and H. Zaragoza. Predictive user click models based on click-through history. In *Proc. CIKM'07*, pages 175–182. ACM, 2007.
[19] F. Radlinski and S. Dumais. Improving personalized web search using result diversification. In *Proc. SIGIR'06*, pages 691–692. ACM, 2006.
[20] F. Radlinski, R. Kleinberg, and T. Joachims. Learning diverse rankings with multi-armed bandits. In *Proc. ICML'08*, pages 784–791. ACM, 2008.
[21] F. Radlinski, M. Szummer, and N. Craswell. Inferring query intent from reformulations and clicks. In *Proc. WWW'10*, pages 1171–1172. ACM, 2010.
[22] D. Rafiei, K. Bharat, and A. Shukla. Diversifying web search results. In *Proc. WWW'10*, pages 781–790. ACM Press, 2010.
[23] E. Sadikov, J. Madhavan, L. Wang, and A. Halevy. Clustering query refinements by user intent. In *Proc. WWW'10*, pages 841–850. ACM, 2010.
[24] R. Santos, C. Macdonald, and I. Ounis. Exploiting query reformulations for web search result diversification. In *Proc. WWW'10*, pages 881–890. ACM Press, 2010.
[25] R. Santos, J. Peng, C. Macdonald, and I. Ounis. Explicit search result diversification through sub-queries. In *Proc. ECIR'10*, pages 87–99. ACM Press, 2010.
[26] R. L. Santos, C. Macdonald, and I. Ounis. Selectively diversifying web search results. In *Proc. CIKM'10*, pages 1179–1188, New York, USA, 2010. ACM.
[27] E. Vee, U. Srivastava, J. Shanmugasundaram, P. Bhat, and S. A. Yahia. Efficient computation of diverse query results. In *Proc. ICDE'08*, pages 228–236. IEEE CS, 2008.
[28] O. Zamir and O. Etzioni. Grouper: a dynamic clustering interface to web search results. In *Proc. WWW'99*, pages 1361–1374. Elsevier North-Holland, Inc., 1999.
[29] C. Zhai. *Risk minimization and language modeling in Information Retrieval*. PhD thesis, CMU, 2002.
[30] C. Zhai and J. Lafferty. A risk minimization framework for information retrieval. *IP&M*, 42(1):31–55, 2006.
[31] C. X. Zhai, W. W. Cohen, and J. Lafferty. Beyond independent relevance: methods and evaluation metrics for subtopic retrieval. In *Proc. SIGIR'03*, pages 699–708. ACM, 2003.
[32] C.-N. Ziegler, S. M. McNee, J. A. Konstan, and G. Lausen. Improving recommendation lists through topic diversification. In *Proc. WWW'05*, pages 22–32. ACM, 2005.

# APPENDIX

## A. OPTSELECT ALGORITHM

The pseudocode of Algorithm 2 describes the steps for solving the MAXUTILITY_DIVERSIFY($k$) problem.

---
**Algorithm 2** OptSelect $(q, S_q, R_q, k)$
---
01. $S \leftarrow \emptyset$;
/* Heap($n$) instantiates a new $n$-size heap */
02. $M \leftarrow new$ Heap($k$);
03. **For Each** $q' \in S_q$ **Do**
04. $\quad M_{q'} \leftarrow new$ Heap($\lfloor k \cdot P(q'|q) \rfloor + 1$);
05. $\quad$ **For Each** $d \in R_q$ **Do**
06. $\quad\quad$ **If** $\widetilde{U}(d|R_{q'}) > 0$ **Then** $M_{q'}.push(d)$; **Else** $M.push(d)$;
07. **For Each** $q' \in S_q$ s.t. $M_{q'} \neq \emptyset$ **Do**
08. $\quad x \leftarrow$ pop $d$ with the max $\widetilde{U}(d|q)$ from $M_{q'}$;
09. $\quad S \leftarrow S \cup \{x\}$;
10. **While** $|S| < k$ **Do**
11. $\quad x \leftarrow$ pop $d$ with the max $\widetilde{U}(d|q)$ from $M$;
12. $\quad S \leftarrow S \cup \{x\}$;
13. **Return** ($S$);
---

## B. DATASETS USED

To assess effectiveness we have followed the guidelines of the Diversity Task of TREC. We have used the ClueWeb-B dataset, i.e. the subset of the TREC ClueWeb09 dataset[4] collection used in the TREC 2009 Web track's Diversity Task, comprising a total of 50 million English Web documents. A total of 50 topics were available for this task. Each topic includes from 3 to 8 sub-topics manually identified by TREC assessors, with relevance judgements provided at subtopic level. As an example the first TREC topic is identified by the query *obama family tree*, and three subtopics are provided: i) *Find the TIME magazine photo essay "Barack Obama's Family Tree"*, ii) *Where did Barack Obama's parents and grandparents come from?*, and iii) *Find biographical information on Barack Obama's mother*.

The two query logs used are AOL and MSN. The AOL data-set contains about 20 millions of queries issued by about 650,000 different users. The queries were submitted to the AOL search portal over a period of three months from 1st March, 2006 to 31st May, 2006. The MSN Search query log contains 15 millions of queries submitted to the MSN US search portal over a period of one month in 2006. Queries are mostly in English. Both query logs come with all the information needed to address the diversification problem according to our approach.

## C. EVALUATION BASED ON QUERY LOG DATA

The second evaluation we propose exploits the user sessions and the query specializations coming from the query logs of two commercial search engines. The aim of this evaluation is to show the importance of having a good diversification method based on real users' interests.

The two query logs were split into two different subsets. The first one (containing approximatively the 70% of the queries) was used for training (i.e., to build the data structures described in the previous section), and the second one for testing. For any ambiguous query $q$ obtained by applying the Algorithm 1 to the test set of each query log, we first submitted the query to the Yahoo! BOSS search engine, then we re-ranked the results list by means of the Algorithm 2 to obtain the corresponding diversified list of results. Finally, we compared the two lists obtained by means of the utility function as in Definition 2. The goal is to show that our diversification technique can provide users with a list of $k$ documents having a utility greater than the top-$k$ results returned by the Yahoo! BOSS Search Engine.

---
[4] http://boston.lti.cs.cmu.edu/Data/clueweb09/

To assess the impact of our diversification strategy on the utility of the diversified results list, we computed the ratio between the normalized utilities of the results in $S$ and the top-$k$ results in $R_q$, i.e., the diversified and the original one. More formally, we computed

$$\frac{\sum_{i=1}^{k} \widetilde{U}(d_i \in S)}{\sum_{i=1}^{k} \widetilde{U}(d_i \in R_q)}$$

where $S$ is the diversified list produced by OptSelect, whereas $R_q$ is the original list of results obtained from Yahoo! BOSS. It is clear that if the two lists share all the results, the ratio is equal to 1.

In our tests, we set the number of results retrieved from Yahoo! BOSS ($|R_q|$) equal to 200, while both $|R_{q'}|$ and $k$ equal to 20.

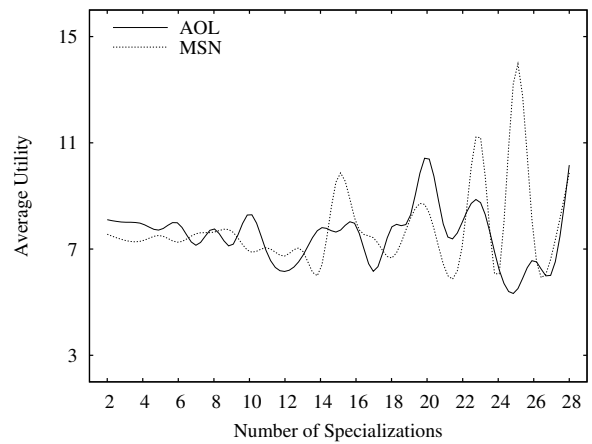

**Figure 1: Average utility per number of specializations referring to the AOL and MSN query logs.**

Figure 1 shows the average utility per number of specializations for the two query logs considered in our experiments. In all cases taken into account, our method diversifies the final list by improving the usefulness measure for a factor ranging from 5 to 10 with respect to the usefulness of the original result set.

Furthermore, we measured the number of times our method is able to provide diversified results when they are actually needed, i.e., a sort of *recall* measure. This was done by considering the number of times a user, after submitting an ambiguous/faceted query, issued a new query that is a specialization of the previous one. In both cases we are able to provide diversified results for a large fraction of the queries. Concerning AOL, we are able to diversify results for the 61% of the cases, whereas for MSN this *recall* measure raises up to 65%.